\documentclass[twocolumn]{aastex6}
\usepackage{natbib}
\usepackage{color}
\usepackage{hyperref} %----set up hyper link to citations, equations...
\def \bea {\begin{eqnarray}}
\def \ena {\end{eqnarray}}               
\def \bee {\begin{equation}}
\def \ene {\end{equation}}
\def    \simlt  {\lower.5ex\hbox{$\; \buildrel < \over \sim \;$}}
\def    \simgt  {\lower.5ex\hbox{$\; \buildrel > \over \sim \;$}}

\newcommand     \mum    {\,\mu{\rm m}}  % to use in math mode

\def	\cm		{\,{\rm {cm}}}

\def	\eV		{\,{\rm {eV}}\,}

\def	\g		{\,{\rm g}}

\def	\K		{\,{\rm K}}

\def	\AU		{\,{\rm {AU}}}
\def	\pc		{\,{\rm {pc}}}

\def	\s		{\,{\rm s}}
\def	\sr		{\,{\rm {sr}}}

\def	\H		{\rm H}

\def    \bB     	{{\bf  B}}
\def    \bE     	{{\bf  E}}
\def    \bv     	{{\bf  v}}

\font\mib=cmmib10
\def\bmu{\hbox{\mib\char"16}}
\def\bGamma{\hbox{\mib\char"00}}

\begin{document}
\shorttitle{Effect of interstellar magnetic fields}
\shortauthors{Hoang and Loeb}
\title{Electromagnetic forces on a relativistic spacecraft in the interstellar medium}
\author{Thiem Hoang}
\affil{Korea Astronomy and Space Science Institute, Daejeon 34055, Korea; thiemhoang@kasi.re.kr}
\affil{Korea University of Science and Technology, Daejeon 34113, Korea}
\author{Abraham Loeb}
\affil{Harvard-Smithsonian Center for Astrophysics, 60 Garden Street, Cambridge, MA, USA; aloeb@cfa.harvard.edu}

\begin{abstract}
A relativistic spacecraft of the type envisioned by the Breakthrough Starshot initiative will inevitably get charged through collisions with interstellar particles and UV photons. Interstellar magnetic fields would, therefore, deflect the trajectory of the spacecraft. We calculate the expected deflection for typical interstellar conditions. We also find that the charge distribution of the spacecraft is asymmetric, producing an electric dipole moment. The interaction between the moving electric dipole and the interstellar magnetic field is found to produce a large torque, which can result in fast {oscillation} of the spacecraft around the axis perpendicular to the direction of motion, with a period of $\sim$ 0.5 hr. We then study the spacecraft rotation arising from impulsive torques by dust bombardment. Finally, we discuss the effect of the spacecraft rotation and suggest several methods to mitigate it. 

\end{abstract}

\section{Introduction}\label{sec:intro}
The Breakthrough Starshot initiative\footnote{https://breakthroughinitiatives.org/Initiative/3} aims to launch gram-scale spacecraft with miniaturized electronic components (such as camera, navigation, and communication systems) to relativistic speeds ($v\sim 0.2c)$. This will enable the spacecraft to reach the nearest stars, like $\alpha$ Centauri (distance of 1.34 pc), within a human lifetime. Such spacecraft would also revolutionize exploration of the solar system, the neighboring Oort cloud, and the local interstellar medium (ISM).

The recent discovery of an Earth-mass planet around the nearest star, Proxima b \citep{AngladaEscude:2016hq}, gives a strong boost to the concept of a flyby mission. The question whether Proxima b hosts an atmosphere will be addressed by the {\it James Webb Space Telescope} \citep{Kreidberg:2016tc}, but a spacecraft could probe more closely signs of life. The main challenge in navigating the spacecraft involves achieving a high accuracy in aiming at the target. 

\cite{2017ApJ...837....5H} studied in detail the damage to a relativistic spacecraft in the ISM. Dust collisions incur the most severe damage, which can be mitigated by adding a shield of a few mm in thickness made of a strong material like graphite. In this paper, we investigate the dynamics of the spacecraft, including the ISM effect on the spacecraft trajectory and orientation. 

The trajectory of a relativistic spacecraft will be affected by both gravitational and non-gravitational forces. In practice, the spacecraft will be subject to drag forces due to collisions with interstellar gas and dust. Moreover, the spacecraft will inevitably get charged by such collisions. As a result, the charged spacecraft will experience a Lorentz force from the interstellar magnetic field and gets deflected from its straight trajectory. We will first quantify the corresponding deflection of the spacecraft trajectory. 

The spacecraft is also expected to experience impulsive torques due to dust bombardment, which induces the spacecraft rotation. Moreover, the resulting charge distribution is likely to be asymmetric, producing an electric dipole moment. The interaction of the moving dipole with the interstellar magnetic field will {turn} the spacecraft around its center of mass, leading to enhanced damage by interstellar matter. {Throughout this paper, we will use the cgs units.}

The structure of the paper is as follows. We first discuss the charging of a relativistic spacecraft by collisions and photoemission, and calculate the maximum resulting surface potential in Section \ref{sec:charg}. In Section \ref{sec:force}, we consider the forces and torques that a charged spacecraft experiences in the magnetized ISM. In Section \ref{sec:def_ISM}, we calculate the resulting deflection of the spacecraft in the interstellar magnetic field. Section \ref{sec:rot} is devoted for studying the oscillation and rotation of the spacecraft by regular and impulsive torques. Finally, we discuss the implications of our results in Section \ref{sec:dis} and summarize our conclusions in Section \ref{sec:sum}.

\section{Charging of a relativistic spacecraft}\label{sec:charg}

For our present study, we use the model Starchip considered in \cite{2017ApJ...837....5H}, in the form of a thin tube of height $H$, width $W$, and length $L$. The optimal shape of the spacecraft with the smallest frontal cross-section is needle-like with $H=W\ll L$, as shown in Figure \ref{fig:model}. We assume that the spacecraft is moving with the velocity $v$ parallel to the long axis, such that the cross-section surface area of the spacecraft is $A_{\rm sf}=WH=W^{2}$.

\subsection{Collisional charging}
\subsubsection{General consideration}
Upon collisions with the spacecraft, energetic electrons and ions penetrate and transfer their kinetic energy to the target electrons. The projectiles are stopped within the spacecraft eventually if their penetration depth is shorter than the spacecraft length \citep{2017ApJ...837....5H}. During their path within the spacecraft, energetic incident electrons can liberate numerous secondary electrons. Some of these electrons can have enough kinetic energy to reach the surface. If their kinetic energy at the surface is above the surface potential, they will exit as free electrons. This electron emission process leads to a positive charge of the spacecraft.  

\begin{figure}
\centering
\includegraphics[width=0.5\textwidth]{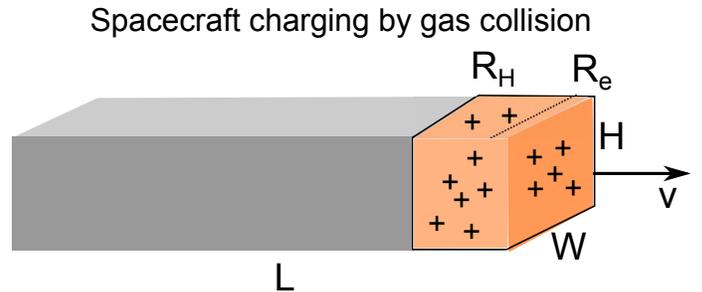}
\caption{Schematic of the needle-like spacecraft of height $H$, width $W$, and length $L$. The spacecraft surface area is $A_{\rm sf}=WH$. A thin surface layer can be highly charged by electron emission from collisions with interstellar electrons and protons, while the sides of the spacecraft are weakly charged through photoelectric emission of electrons by the absorption of UV photons.}
\label{fig:model}
\end{figure}

Figure \ref{fig:model} shows a sketch of the spacecraft charging process through which a thin surface layer becomes positively charged through collisions with electrons and protons. The volume of this charged region (colored) is constrained by the penetration depth of protons $R_{\H}$, much larger then the electron penetration depth $R_{e}$. We define $a_{\rm eff}$ as the radius of the equivalent sphere that has the same volume as the charged region,
\bea
a_{\rm eff}\equiv \left(\frac{3R_{\H}A_{\rm sf}}{4\pi}\right)^{1/3}=\left(\frac{3R_{\H}W^{2}}{4\pi}\right)^{1/3}.\label{eq:a_sf}
\ena
The charged region and $a_{\rm eff}$ vary with $v$ because the penetration depth $R_{\H}$ is a function of the spacecraft speed $v$ (see \citealt{2017ApJ...837....5H}).

\subsubsection{Electron bombardment}
Secondary electron emission by electron bombardment is well studied in the literature. The interested reader is referred to the classical review by \cite{Hachenberg:1959cl} and a more recent review by \cite{1980PhRvB..22.2141S}, for more details. Here, we only provide the essential formulae.

Let $\delta(E_{0},0)$ be the electron emission yield, which is defined by the total number of electrons emitted per an incident electron of initial energy $E_{0}$. For a bulk solid, the emission yield can be described by:
\bea
\delta (E_{0},0)= \delta_{m}\frac{4E_{0}/E_{m}}{(1+E_{0}/E_{m})^{2}},\label{eq:deltaE0}
\ena
where $E_{m}$ and $\delta_{m}$ are the peak energy and maximum yield (\citealt{Hachenberg:1959cl}; \citealt{1979ApJ...231...77D}). Experimental data show that, for quartz, $E_{m}=420 \eV$ and $\delta_{m}=2.9$. For graphite, $E_{m}=250 \eV$ and $\delta_{m}=1$ (\citealt{Hachenberg:1959cl}).

For $E_{0}\gg E_{m}$, Equation (\ref{eq:deltaE0}) scales as $1/E_{0}$, as expected from the Rutherford cross-section. For $E_{0}\ll E_{m}$, the yield is linearly dependent on $E_{0}$.

The energy distribution of secondary electrons is uncertain. The emission yield for the secondary electron with energy  $E>E_{s}$ is approximately given by (\citealt{1979ApJ...231...77D}):
\bea
\delta(E_{0},E)=\frac{\delta(E_{0},0)}{[1+0.125(E/\rm eV)^{2}]^{1/2}},\label{eq:delta_E}
\ena
where $E_{s}$ is the threshold for escaping from the surface, and $E$ is in unit of $\eV$. This provides a good fit to experimental data in solids, and is proportional to $1/E$ for large $E$ as expected from Rutherford scattering of secondary electrons.

The emission yield from the spacecraft of surface electric potential $U$ is given by,
\bea
\delta(E_{0},U) = \delta(E_{0},E= eU),
\ena
where $E_{s}=eU$ is the threshold for free electrons. For our case, the spacecraft is rapidly charged to $eU\gg 1\eV$. In this regime, we can approximate Equation (\ref{eq:delta_E}) as follows:
\bea
\delta(E_{0},U) \approx \frac{\delta(E_{0},0)}{\sqrt{0.125}(E/\eV)}\approx \frac{\delta(E_{0},0)}{0.35(eU/\eV)}.\label{eq:deltaU}
\ena

The left panel of Figure \ref{fig:yield} shows the emission yield versus the spacecraft speed $v$ for different values of the spacecraft potential between $U=0$ and $U=U_{\max,e}/2$. Because $U_{\max,e}$ is a function of $v$ (see Section \ref{sec:max}), we find a steeper decline of the yield with $v$ for increased $U$.

\subsubsection{Proton bombardment}
Ion bombardment can also induce the emission of secondary electrons. The physics is similar to electron impact. \cite{Sternglass:1957vr} first presented a theoretical study on electron emission by energetic protons. An analysis for spacecraft under astrophysical conditions is compiled in \cite{1981RPPh...44.1197W}.

Let $\gamma(E_{0},0)$ be the total electron emission yield induced by an proton of initial energy $E_{0}$. For metals, \cite{1992STMP..123....1H} showed that the total emission yield by proton bombardment is proportional to the stopping power $dE/dx$,
\bea
\gamma(E_{0},0) = \Lambda_{e}\frac{dE}{dx},\label{eq:gamma_E0}
\ena
where $\Lambda_{e}$ is specific yield that depends on the material \citep{1980PhRvB..22.2141S}. A good fit for the emission yield by impinging protons shows an independence of $\Lambda_{e}$ on the ion energy $\Lambda_{e} \approx 0.1$ (\AA/eV).

Based on experiments for 17 different projectile ions on carbon targets, \cite{1993PhRvB..48.6832C} found a good fit to their data with $\Lambda_{e} \sim 0.36$ (\AA/eV). 

As in the case of impinging electrons, the emission yield by proton impacts producing secondary electrons of kinetic energy $E> E_{s}$ is given by
\bea
\gamma(E_{0},E) = \frac{\gamma(E_{0},0)}{\left[1+ 0.125(E/\rm eV)^{2}\right]^{1/2}}.\label{eq:gamma_E}
\ena  

Therefore, the emission yield that enters Equation (\ref{eq:gamma_E}) is given by
\bea
\gamma(E_{0},U) = \gamma(E_{0},E= eU),\label{eq:gamma_EU}
\ena
where $E_{s}=eU$ is the threshold for free electrons.

\begin{figure*}
\includegraphics[width=0.5\textwidth]{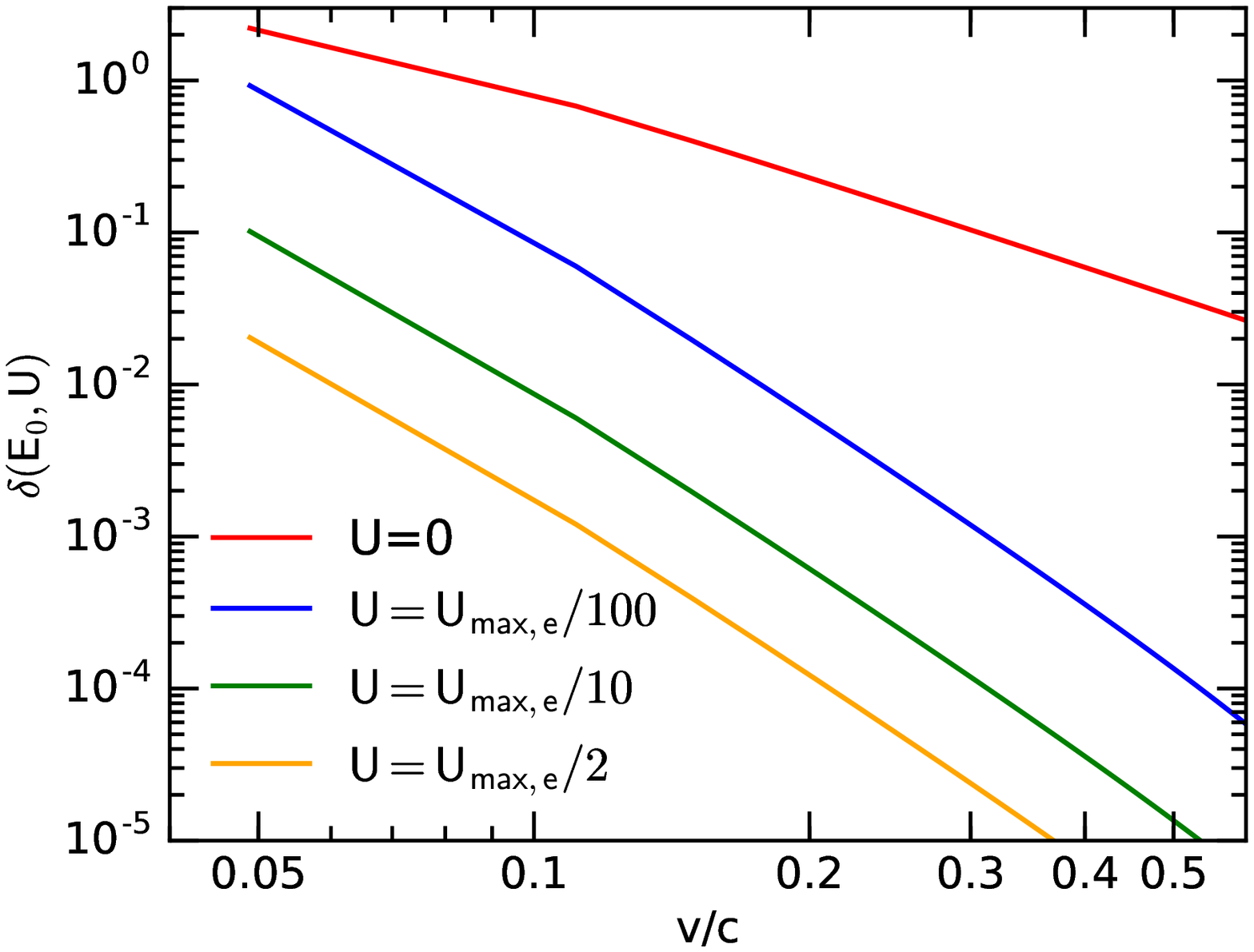}
\includegraphics[width=0.5\textwidth]{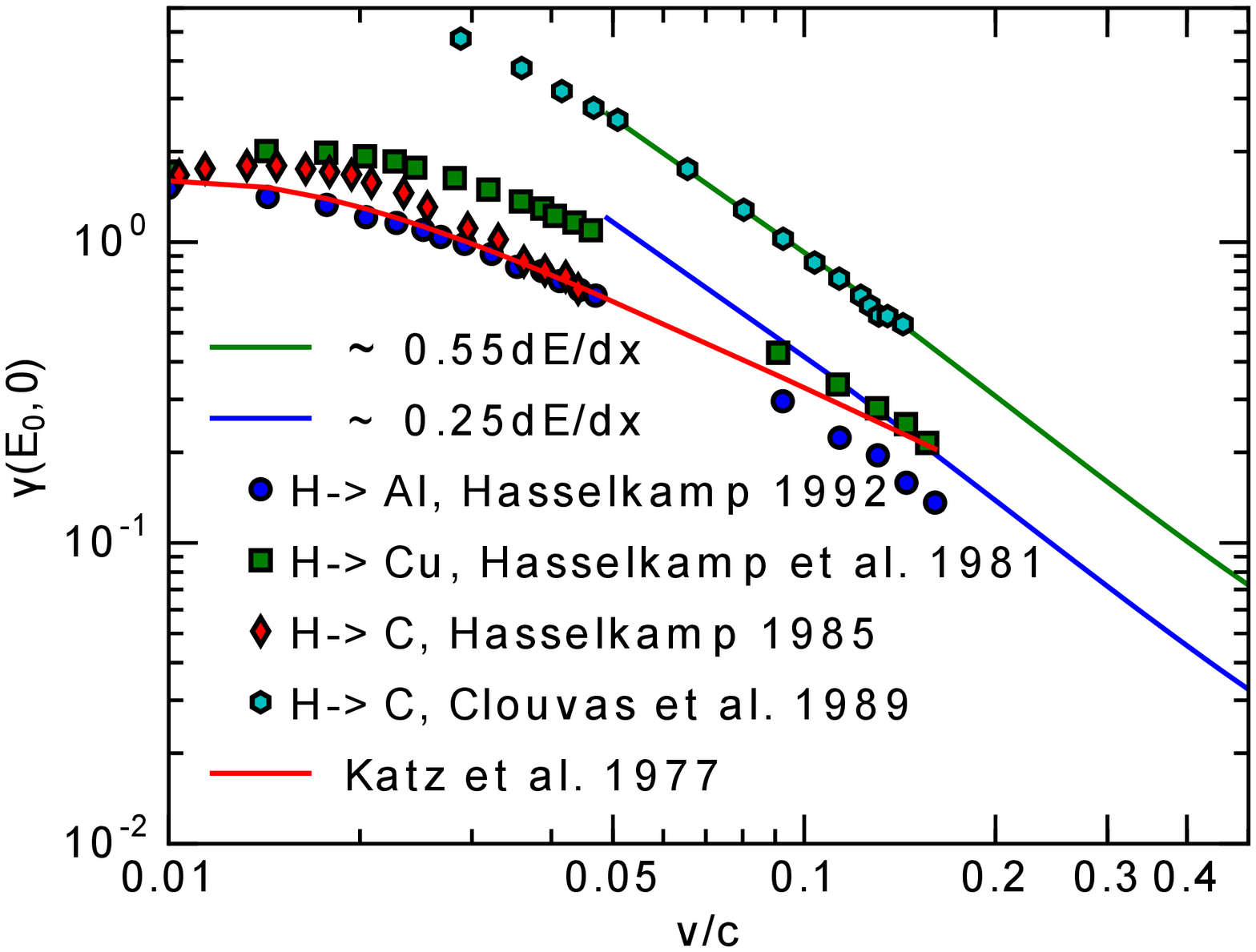}
\caption{{\bf Left panel}: Electron emission yield $\delta(E_{0},U)$ calculated for the different potentials of a graphite spacecraft. Larger $U$ values substantially reduce the emission yield. {\bf Right panel}: Emission yield by proton impact $\gamma(E_{0},0)$ computed vs. experimental data for metals and carbon targets. The fit for the carbon target is calculated using $dE/dx$ of graphite.}
\label{fig:yield}
\end{figure*}

Figure \ref{fig:yield} (right panel) shows $\delta(E_{0},0)$ computed based on Equation (\ref{eq:gamma_E0}) for graphite material in comparison to experimental data from \cite{1989PhRvB..39.6316C} and \cite{1992STMP..123....1H}. Data for metal targets from \cite{1981NucIM.180..349H} are also shown. The stopping power $dE/dx$ for the selected material is taken from \cite{2017ApJ...837....5H}. An excellent agreement between the theory and data is achieved for $v>0.05c$. The analytical fit suggested by \cite{Katz:1977tz} (see also \citealt{1981RPPh...44.1197W}) is not valid for the relativistic case of $v>0.1c$ (see red line). 

For our calculations in the present paper, we calculate $\gamma(E_{0},0)$ using Equation (\ref{eq:gamma_E0}) with $\Lambda_{e}=0.55/2$ (\AA/eV) which provides the best-fit to the data by \cite{1989PhRvB..39.6316C} (see the right panel of Figure \ref{fig:yield}). Here the factor 2 accounts for the fact that the experiments studied by \cite{1989PhRvB..39.6316C} were conducted for thin foils where electrons can escape from both sides. For our long spacecraft, the yield is expected to be a factor 2 smaller because electrons only escape from the front side.

%For the different materials, the parameter $\Lambda_{e}$ spans from 0.1-0.5. Due to the lack of experimental data for quartz, we adopt $\Lambda_{e}=0.5$.

\subsubsection{Heavy ion bombardment}
Energetic heavy ions are expected to produce a larger electron emission yield than protons due to the larger stopping power $dE/dx$. The experiments discussed in \cite{Borovsky:1991wc} and \cite{1993PhRvB..48.6832C} show that the emission yield by heavy ions is about two orders of magnitude larger than that of protons. However, since the interstellar abundance of heavy atoms is more than four orders of magnitude lower than hydrogen, the net emission yield by heavy ions is negligible compared to impinging electrons and protons. Thus, charging by electron emission due to heavy ions can be neglected.

Moreover, bombardment of fast heavy ions can eject target nuclei through a process so-called electronic sputtering \citep{2015ApJ...806..255H}. For our high energy case, this process is shown to be inefficient and yields a rather low sputtering yield. For instance, \cite{2015ApJ...806..255H} found that iron ions can produce a sputtering yield of $Y_{\rm Fe}<0.1$ for $v>0.1c$. The net yield for Fe abundance of $X_{\rm Fe}\sim 10^{-5}$ is $X_{\rm Fe}Y_{\rm Fe}< 10^{-6}$, which is negligible compared to the electron emission by electrons and protons.

\subsubsection{Maximum surface electric potential and Coulomb explosions}\label{sec:max}
The accumulation of positive charge by collisions gradually increases the electric potential of the spacecraft surface $U$. When the potential is sufficiently large such that the liberated electrons cannot overcome the surface potential to exit from it as free electrons, the spacecraft is maximally charged.

The maximum surface potential $U_{\max,p}$ achieved by impinging particles $p$ is determined by the maximum energy that the projectile can transfer to the target atomic electron:
\bea
eU_{\rm max,p}= T_{\rm max,p}, \label{eq:Umax}
\ena
which yields $eU_{\max,e}=m_{e}v^{2}/2$ for impinging electrons and $eU_{\max,\H}=2m_{e}v^{2}=4eU_{\max,e}$ for impinging protons (see \citealt{2015ApJ...806..255H}). 

The corresponding maximum charge is equal to
\bea
Z_{\rm sp,max}= \frac{U_{\rm max,p}a_{\rm eff}}{e},\label{eq:Zmax}
\ena
where $a_{\rm eff}$ is the radius of the charged region that also depends on the spacecraft speed $v$ (see Figure \ref{fig:model}).

We note that $U_{\max,\H}$ is much smaller than the Coulomb disruption limit, $e\phi_{\rm Coul,\max}\simeq 10^{4}\left(S_{\max}/10^{10}{\rm dyn}\cm^{-2} \right)^{1/2}(a_{\rm eff}/0.1\cm)$ keV. Here $S_{\max}$ is the maximum tensile strength of the material that can reach $S_{\max}\sim 10^{11}\rm dyn\cm^{-2}$ for ideal material (see \citealt{2015ApJ...806..255H}). Thus, spacecraft damage by Coulomb explosions can be neglected. 

%The corresponding maximum charge is equal to $eZ_{sp,max}/a_{\rm sf}=U_{rm,max}$.

\subsection{Photoelectric Emission}
The spacecraft can be charged by absorption of interstellar UV photons via the photoelectric effect.

The flux of the interstellar UV photons with energy $E$ from $5-13.6\eV$ is \citep{1978ApJS...36..595D}:
\bea
F(E)& =& 1.658\times 10^{6} E -2.152\times 10^{5} E^{2} \nonumber\\
&&+ 6.919\times 10^{3} E^{3} \cm^{-2} s^{-1} \sr^{-1} \eV^{-1},\label{eq:fUV}
\ena
where $E$ is given in units of eV.

The total number density of UV photons is equal to $n_{\rm UV}=\int_{5\eV}^{13.6\eV} EF(E) dE/c= 0.06 \cm^{-3}$. Thus, the UV photon density is an order of magnitude lower than the total electron and proton density in the local ISM of $\sim 0.5\cm^{-3}$. Moreover, the UV photon energy is low ($<20 \eV$), so that liberated electrons cannot have sufficient energy to escape from the front end after the spacecraft is already charged to a sufficiently large potential of $U> 20$ eV. The photoelectric effect is negligible for charging the front side of a relativistic spacecraft compared to gas particle collisions. On the other hand, the photoelectric effect is important for charging the back end and the sides of the spacecraft. Nevertheless, the maximum surface potential achieved by UV photons is much smaller than the maximum potential of the front side of $U_{\rm max,e}\sim 10^{4}$ eV.

\subsection{Electric dipole}
As shown in Figure \ref{fig:model}, a thin surface layer of thickness $\sim R_{\H}$ will be positively charged. The center of charge, $Z_{\rm sp}$, is displaced from the center of mass by $\sim L/2$, producing an electric dipole moment. The electric dipole is directed along the long axis $\hat{\bf x}$ and equal to
\bea
\bmu_{ed} \approx \frac{eZ_{\rm sp}L}{2}~\hat{\bf x}.\label{eq:dipole}
\ena

\section{Forces and Torques Acting on a Charged spacecraft}\label{sec:force}
Below, we will discuss the forces and torques that a charged spacecraft experiences during its journey in the magnetized ISM.

\subsection{Electromagnetic forces}
The charged spacecraft experiences the Lorentz force due to the interstellar magnetic field:
\bea
F_{B} = \frac{eZ_{\rm sp}vB_{\perp}}{c},
\ena
where $B_{\perp}$ is the magnetic field component perpendicular to the direction of motion of the spacecraft.

For the short period when the spacecraft is moving through the solar system, the motion of the solar system with respect to the ambient magnetic field induces an electric field $\bE=\bv_{\odot}\times \bB$. Thus, the electric force is
\bea
F_{E} = eZ_{\rm sp}E= \frac{eZ_{\rm sp}v_{\odot}B_{\perp}}{c}.\label{eq:fE}
\ena
Since $v_{\odot}$ is much lower than $v$, $F_{E}\ll F_{B}$, so this force can be neglected.

{The spacecraft is also subject to the force exerted by the solar wind, which is given by Equation (\ref{eq:fE}) with $v_{\odot}$ replaced by the solar wind velocity $v_{\rm sw}$. Since $v_{\rm sw}\ll c$, the resulting force is negligible.}

\subsection{Drag forces}
Gas atoms that collide with the spacecraft produce a drag force,
\bea
F_{\rm drag,coll} \simeq 1.4n_{\H}m_{\H}v^{2} A_{\rm sf},\label{eq:Fdrag_gas}
\ena
where the factor of $1.4$ accounts for the He with abundance of $10\%$, and $n_{\H}, m_{\H}$ are the proton number density and mass, respectively.

A charged body moving in a plasma is also expected to experience a Coulomb drag force. However, in the relativistic limit of $v\gg (2kT_{\rm gas}/m_{\H})^{1/2}\sim 10^{6}(T_{\rm gas}/10^{4}\K)^{1/2}\cm/\s$. When the spacecraft motion is much faster than the thermal speed of protons, the Coulomb drag is negligible.

The resulting slowing down of the spacecraft by drag forces in the ISM is described the fractional velocity change, $\Delta v/v\sim 2.3\times 10^{-6}(N_{\H}/10^{18}\cm^{-2})(A_{\rm sf}/1\cm^{2})(1\g/M_{\rm sp})$. This implies that the spacecraft will be delayed in reaching the target by about 2 hr over the travel time of 20 years to $\alpha$ Centauri. This finding is consistent with the conclusion obtained in \cite{2016arXiv160401356L}. 

{When the lightsail is open, an additional drag from from interstellar gas is present. However, such drag force is negligible for relativistic thin lightsails \citep{Hoang:2017wz}}.

\subsection{Torques}
\subsubsection{Regular torque on the electric dipole}
A charged spacecraft moving through the interstellar magnetic field is subject to an {\it equivalent} electric field:
\bea
{\bE}' = \frac{\bv\times \bB}{c}.
\ena

\begin{figure}
\includegraphics[width=0.5\textwidth]{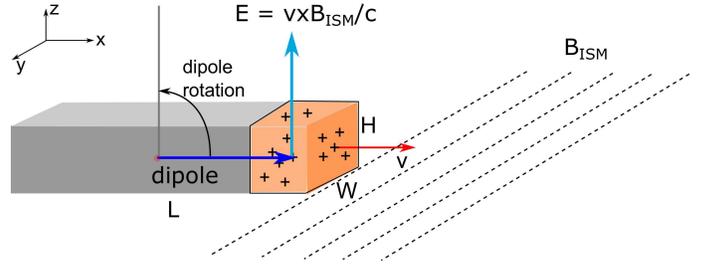}
\caption{Interaction of a charged spacecraft with a velocity $v$ along the x-direction in the interstellar magnetic field along the y-direction. The {equivalent} electric field in the spacecraft is oriented along the z-direction. The torque acting on the dipole results in the oscillation of the spacecraft around the y-axis.}
\label{fig:model_dipole}
\end{figure}

The interaction of the electric dipole with the {equivalent} electric field induces a torque:
\bea
\bGamma_{\rm ed} = \bmu_{\rm ed}\times {\bE}' =\frac{eZ_{\rm sp}L}{2}\left[\hat{\bf x}\times\frac{\bv\times \bB}{c}\right].\label{eq:torq_ed}
\ena
The resulting effect of this torque is to rotate the spacecraft around the $y$-axis, perpendicular to the direction of motion (see Figure \ref{fig:model_dipole}). 

\subsubsection{Impulsive torques by dust impact}
In addition to drag forces, the spacecraft will experience impulsive torques due to collisions with gas and dust. For an axisymmetric surface, gas bombardment results in a zero averaged torque, but an irregularly shaped spacecraft is subject to a net mechanical torque (\citealt{2007ApJ...669L..77L}; \citealt{Hoang:2017tl}). Here we assume that the spacecraft is axisymmetric, such that the regular mechanical torque is zero.

Due to its large mass, a single collision with a dust grain of mass $m_{\rm gr}$ can deposit a considerable angular momentum on the spacecraft over a very short interaction time:
\bea
\delta J \sim \frac{m_{\rm gr}vW}{2}.\label{eq:dL}
\ena

The net increase of the spacecraft angular momentum after experiencing $N_{\rm coll}$ collisions can be estimated using the random walk formula:
\bea
\Delta J = N_{\rm coll}^{1/2}\delta J.\label{eq:deltaJ}
\ena

Assuming that all interstellar dust mass is concentrated in a sizebin of $0.1\mum$ and the dust-to-gas mass ratio of 0.01, the total number of collisions with $0.1\mum$ grains is
\bea
N_{\rm coll} &\sim & \frac{0.01N_{\H}m_{\H}A_{\rm sf}}{(\rho 4\pi/3)\times 10^{-15}\cm^{3}}\nonumber\\
&\sim& 10^{6}\left(\frac{N_{\H}}{10^{18}\cm^{-2}}\right)\left(\frac{A_{\rm sf}}{1\cm^{2}}\right)\left(\frac{3\g\cm^{-3}}{\rho}\right),~~~\label{eq:Ncoll}
\ena
where $N_{\H}$ is the gas column swept by the spacecraft. 

The expected value of N$_{\H}$ to $\alpha$ Centauri is in the range of $3\times 10^{17}-10^{18}\cm^{-2}$ \citep{1996ApJ...463..254L}. Thus, we expect to have $N_{\rm coll}\sim 10^{6}$ collisions by the time the spacecraft reaches $\alpha$ Centauri.

\section{Deflection of spacecraft in the interstellar magnetic field}\label{sec:def_ISM}
Next, we aim to quantify the deflection of the charged spacecraft in the interstellar magnetic field.

\subsection{Gas column required for maximum charge}
Because the charge of the spacecraft increases with time, it is of interest to know when the charge reaches its equilibrium value.

The charging rate by collisions with electrons for a spacecraft of charge $Z$ (potential $U$) is given by,
\bea
J_{Z,e} = -n_{e}vA_{\rm sf}\left(s_{e}-\delta \right),
\ena
where $s_{e}$ is the sticking coefficient, and $\delta=\delta(E_{0},U)$ with $E_{0}=m_{e}v^{2}/2$.

Similarly, the charging rate due to proton bombardment is equal to,
\bea
J_{Z,i} = n_{\H}vA_{\rm sf}\left(s_{i} + \gamma\right),\label{eq:JZ_ion}
\ena
where $s_{i}$ is the sticking coefficient, and $\gamma=\gamma(E_{0},U)$ with $E_{0}=m_{\H}v^{2}/2$. Here we disregard the possibility of recombination of protons with target electrons because the protons move with much higher speeds than the Bohr velocity. The focusing effect is also negligible because the maximum energy potential is always much smaller than the proton kinetic energy. For an elongated design, the sticking coefficients, $s_{e}=s_{i}=1$.

\begin{figure}
\includegraphics[width=0.5\textwidth]{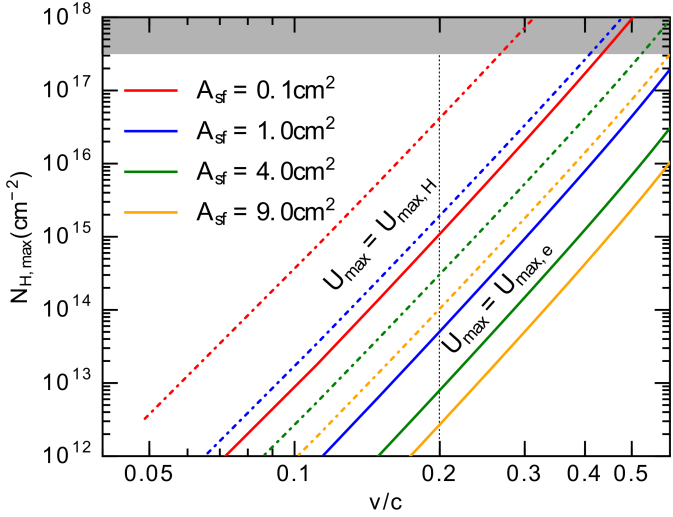}
\caption{Gas column density required for the spacecraft to reach its maximum electric potential $U_{\rm max,e}$ (solid lines) and $U_{\max,\H}$ (dotted lines) for different surface areas $A_{\rm sf}$. Shaded area marks the expected gas column toward $\alpha$ Centauri \citep{1996ApJ...463..254L}, and the vertical line indicates the typical velocity of $v=0.2c$.}
\label{fig:NHmax}
\end{figure}

The increase in the spacecraft electric potential due to collisional charging is given by,
\bea
\frac{dU}{dt} = A_{\rm sf}n_{\H}v\frac{e^{2}}{a_{\rm eff}} \left[\delta(v,U)+\gamma(v,U)\right].\label{eq:dUdt}
\ena

Thus, the gas column swept by the spacecraft to reach the maximum electric potential, denoted by $N_{\rm H,max}$, is calculated as,
\bea
N_{\H,\max}A_{\rm sf}\frac{e^{2}}{a_{\rm eff}} & =& \int_{0}^{U_{\rm max,e}} \frac{dU}{\delta +\gamma }+\int_{U_{\max,e}}^{U_{\max,H}} \frac{dU}{\gamma}\label{eq:NHmax}\\
&\simeq& 0.35\left[\frac{(eU_{\max,e}/\eV)^{2}}{\delta_{0}+\gamma_{0}} +\frac{15(eU_{\max,e}/\eV)^{2}}{\gamma_{0}}\right]\rm eV,\nonumber
\ena
where $\delta(v,U)$ was substituted from Equation (\ref{eq:deltaU}), $\delta_{0}=\delta(E_{0},0)$, $\gamma_{0}=\gamma(E_{0},0)$, $eU_{\max,e}$ is in units of eV, and $a_{\rm eff}$ is given by Equation (\ref{eq:a_sf}).

The first and second terms in Equation (\ref{eq:NHmax}) determines the maximum column required to reach $U_{\max,e}$ and $U_{\max,\H}$, respectively. 

Note that the maximum surface potential is only determined by the energy transfer $T_{\max}$, which is independent of the spacecraft material, leading to the weak dependence of $N_{\H,\max}$ on the material. Therefore, in the following, we present our numerical results for a spacecraft made of graphite only.

Figure \ref{fig:NHmax} shows the column density of gas required to produce the maximum potential $U_{\max,e}$ (solid lines) and $U_{\max,\H}$ (dotted lines). The latter is much larger, as expected from Equation (\ref{eq:NHmax}). The increase of $N_{\rm H,max}$ is expected due to the increase of $U_{\max}$ with $v$ as given by Equation (\ref{eq:Umax}). For a speed of $v=0.2c$, the spacecraft needs to traverse a gas column of $N_{\H}<10^{15}\cm^{-2}$ in order to reach $U_{\rm max,e}$, which corresponds to less than $0.1\%$ of the pathlength to $\alpha$ Centauri. {A gas column of 16 times larger is required to reach $U_{\max,\H}$. In what follows, we adopt $U=U_{\max,e}$ and $Z_{\rm sp}=Z_{\rm sp,\max}$ in our numerical calculations. Results for taking $U=U_{\max, \H}$ will also be discussed.} 

\subsection{Uniform interstellar magnetic field}

The charged spacecraft would move in a helical trajectory in {a uniform magnetic field}. The gyroradius of a charged spacecraft
\bea
R_{g} = \frac{M_{\rm sp}vc}{eZ_{\rm sp}B_{\perp}},\label{eq:Rg}
\ena
provides the curvature of the spacecraft trajectory in the magnetic field. 

The deflection impact parameter from the target at a distance $D\ll R_{g}$ is given by,
\bea
b= D\alpha \approx \frac{D^{2}}{R_{g}}= \frac{D^{2}eZ_{\rm sp}B_{\perp}}{M_{\rm sp}vc},
\ena
where $\sin\alpha \sim\alpha\sim D/R_{g}$. The above equation can be rewritten as
\bea
\left(\frac{b_{\AU}}{D_{\pc}^{2}B_{\mu G}}\right) \simeq 0.017 \left(\frac{Z_{\rm sp}}{10^{10}}\right)\left(\frac{1\g}{M_{\rm sp}}\right)\left(\frac{0.2c}{v}\right),\label{eq:bdefl}
\ena
where $b_{\AU}=(b/1\AU), D_{\pc}=(D/1\pc)$, and $B_{\mu G}=(B_{\perp}/1\mu$G). 

The magnetic field of the local ISM near the Sun is studied in \cite{2012ApJ...760..106F}. The mean magnetic strength of $\sim 3\mu$G is inferred from observations by {\it Interstellar Boundary Explorer} (IBEX) in \cite{2016ApJ...818L..18Z}. The strength of interstellar magnetic field toward $\alpha$ Centauri is uncertain, but likely between $3-10\mu$G. Thus, the corresponding deflection distance is $b\sim 0.1-0.3\AU$ for the magnetic field in this range.

\begin{figure}
\includegraphics[width=0.5\textwidth]{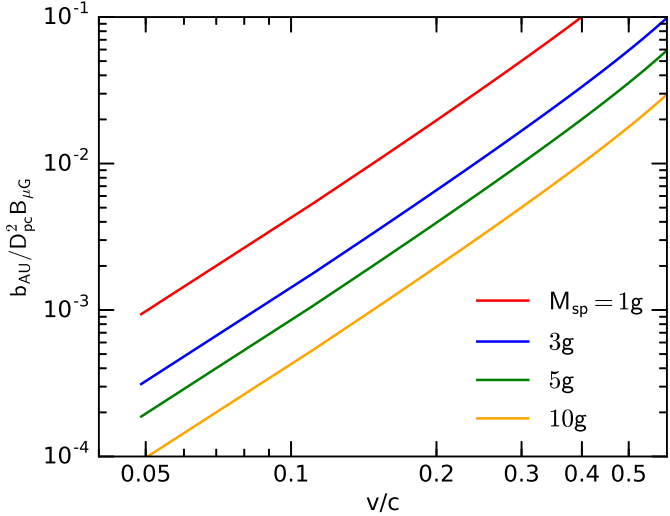}
\caption{Deflection distance off the intended target planet versus spacecraft speed for the different spacecraft masses. Larger speeds induce larger deflections due to the increase of the maximum surface charge with the speed as $v^{2}$. }
\label{fig:dfl_uniform}
\end{figure}

Figure \ref{fig:dfl_uniform} shows the deflection distance ($b/D^{2}B$) as a function of $v$ for the different spacecraft masses, calculated for $Z_{\rm sp}=Z_{\rm sp,\max}$. The deflection distance increases with $v$ but decreases with the spacecraft mass. 

{We note that after the spacecraft is charged to $U_{\max,\H}$, the deflection impact factor is increased by a factor of $U_{\rm max,H}/U_{\rm max,e}=4$.}

\subsection{Effect of non-uniform interstellar medium}
The actual ISM is not uniform, with fluctuating gas density, which results in the fluctuation of the collisional charging. However, after the maximum charge is obtained, the effect of density fluctuation is rather weak. The chance for the presence of dense clumps that can heat the spacecraft to melting temperature is low (see \citep{2017ApJ...837....5H}). Therefore, we consider a uniform ISM for simplicity.

\subsection{Deflection by stellar magnetic fields}

There is no information about the magnetic field in the astrosphere of $\alpha$ Centauri A and B. However, similarly to the Sun, we expect the stellar winds to create magnetic fields out to a distance of $\sim 100$ AU. The magnetic field may be locally amplified due to a bow-shock around the stellar wind. {The magnetic field and stellar wind around Proxima Centauri were studied in \cite{2016ApJ...833L...4G}.} However, the path within the stellar neighborhood is rather short. As a result, the deflection is negligible compared to the deflection caused by the interstellar magnetic field. It is worth noting that, due to the presence of the stellar wind, the atmosphere of Proxima b may be detected through its radio aurora emission \citep{Burkhart:2017wn}.

\section{Spacecraft Oscillation and Rotation}\label{sec:rot}
\subsection{Oscillation by equivalent electric field}
The oscillation of the spacecraft due to the torque (see Equation \ref{eq:torq_ed}) on the moving dipole in the interstellar magnetic field is governed by the equation of motion:
\bea
\frac{I_{\perp}d\omega}{dt} = \frac{eZ_{\rm sp}L}{2c}vB_{\perp}\sin\phi,\label{eq:dJdt_ed}
\ena
where $I_{\perp}$ is the moment of inertia for rotation along the axis perpendicular to the direction of motion, and $\phi$ is the angle between the dipole and $\bE'$ (see Figure \ref{fig:model_dipole}). 

The inertia moments for the rotation parallel and perpendicular to the long axis are $I_{\|}=M_{\rm sp}W^{2}/6$ and $I_{\perp}=M_{\rm sp}(W^{2}+L^{2})/12$, with $M_{\rm sp}=\rho L W^{2}$ and $\rho$ being the mass density.

Adopting the small angle approximation with $\omega = -d\phi/dt$, Equation (\ref{eq:dJdt_ed}) yields the oscillation period of the dipole with respect to $\bE'$:
\bea
T_{\rm ed}&=&\frac{2\pi}{\Omega} = 2\pi\left(\frac{2I_{\perp}c}{eZ_{\rm sp}LvB_{\perp}}\right)^{1/2}\nonumber\\
&\simeq& 0.37 \frac{(M_{\rm sp}/1\g)}{(5\cm/L)}\left(\frac{10^{10}}{Z_{\rm sp}}\right)\frac{(0.2c/v)}{(B_{\perp}/5\mu G)} ~~\rm hr,\label{eq:Trot}
\ena
which implies that the spacecraft will spin rapidly around the axis perpendicular to the direction of motion on a period of $\sim 0.5$ hr. {When the spacecraft is charged to $U_{\max,\H}$, the oscillation rate ($1/T_{\rm ed}$) is increased by a factor of $U_{\rm max,\H}/U_{\rm max,e}=4$.}

\begin{figure}
\includegraphics[width=0.5\textwidth]{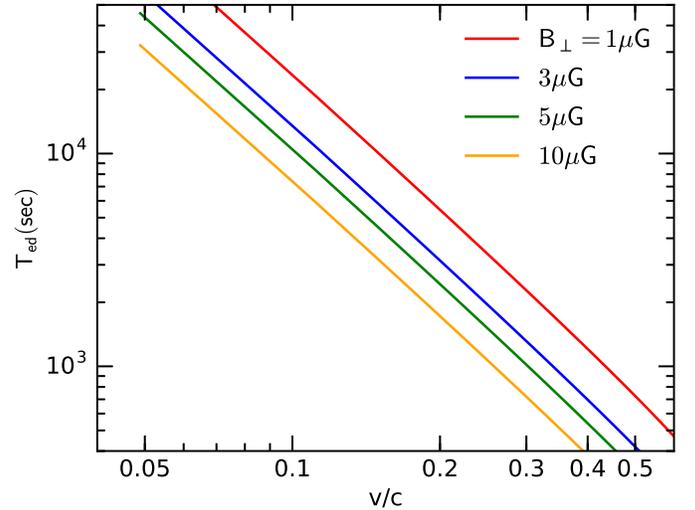}
\caption{{Oscillation} period of the spacecraft $T_{\rm ed}$ vs. its speed for the different magnetic field strengths, assuming spacecraft dimensions $L=5\cm, W=0.3\cm$ and mass $M_{\rm sp}=1\g$ are chosen.}
\label{fig:Trot}
\end{figure}

Figure \ref{fig:Trot} shows $T_{\rm ed}$ calculated from Equation (\ref{eq:Trot}) for different values of $B_{\perp}$. The spacecraft will {oscillate} faster when moving at higher speeds $v$ or traverses a region with an enhanced interstellar magnetic field. 

{Upon entering the stellar magnetosphere, the strong stellar magnetic field can substantially enhance the rate of spacecraft oscillation through its electric dipole as shown in Figure \ref{fig:Trot} by a factor of $B/5\mu G$.}

\subsection{Rotation by dust bombardment}

Using $\delta J$ from Equation (\ref{eq:dL}), we can estimate the angular velocity arising from a single collision with a dust grain of mass $m_{\rm gr}$:
\bea
\delta \omega &=& \frac{\delta J}{I_{\perp}} \approx \frac{6m_{\rm gr}vW}{M_{\rm sp}(L^{2}+W^{2})}\\
&\simeq& 10^{-6}a_{-5}^{3}\left(\frac{v}{0.2c}\right)\frac{(W/0.3\cm)}{M_{\rm sp}(L/5\cm)^{2}}\s^{-1},~~~~\label{eq:domega}
\ena
where $a_{-5}=(a_{\rm gr}/10^{-5}\cm)$. Thus, a single collision can set the spacecraft to spin with $\delta\omega\sim 10^{-6}\s^{-1}$.

For the case in which all dust mass is concentrated in a single size $a_{\rm gr}$ (i.e., single size distribution), the rotation frequency after $N_{\rm coll}$ (Eq. \ref{eq:Ncoll}) collisions is calculated using the total angular momentum $\Delta J$ from Equation (\ref{eq:deltaJ}):
\bea
\Omega &\sim& \frac{\Delta J}{I_{\perp}}\sim \sqrt{N_{\rm coll}}\delta \omega\nonumber\\
&\simeq& 10^{-3}\left(\frac{N_{\H}}{10^{18}\cm^{-2}}\right)\frac{a_{-5}^{3}}{(0.2c/v)}\frac{(W/0.3\cm)}{M_{\rm sp}(L/5\cm)^{2}}\s^{-1}.~~~~~
\ena
This yields the rotation period:
\bea
T_{\rm coll}&=&\frac{2\pi}{\Omega}\nonumber\\
&\simeq& 1.7\left(\frac{10^{18}\cm^{-2}}{N_{\H}}\right)\frac{(0.2c/v)}{a_{-5}^{3}}\frac{M_{\rm sp}(L/5\cm)^{2}}{(W/0.3\cm)} \rm hr,~~~~~\label{eq:Tcoll}
\ena
which corresponds to one revolution over a pathlength of $2\AU$ at $v=0.2c$.

To consider the effect of spacecraft collisions with grains of different sizes, let adopt the grain size distribution from \cite{Mathis:1977p3072}: $f(a)\equiv n_{\H}^{-1}dn/da=A_{\rm MRN}a^{-3.5}$ with $A_{\rm MRN}=10^{-25.16}\cm^{-2.5}$. By using the random walk formula, we obtain 
\bea
\Omega^{2}&=&\int_{a_{\min}}^{a_{\max}} \left(\frac{6vW}{M_{\rm sp}(L^{2}+W^{2})}\right)^{2} N_{\H} m_{\rm gr}^{2}(a) f(a) da\nonumber\\
&=&  \left(\frac{4\pi}{3}\times\frac{6m_{\rm gr}vW}{M_{\rm sp}(L^{2}+W^{2})}\right)^{2} N_{\H} A_{\rm MRN}\frac{a_{\max}^{3.5}}{3.5},\label{eq:Omega2}
\ena
where $a_{\min}$ and $a_{\max}$ are the lower and upper cutoff of the size distribution. With the standard cutoff of the interstellar dust $a_{\max}=0.25\mum$ (\citealt{Mathis:1977p3072}), we obtain
which yields
\bea
T_{\rm coll}\simeq 0.78\left(\frac{10^{18}\cm^{-2}}{N_{\H}}\right)\left(\frac{0.2c}{v}\right)\frac{M_{\rm sp}(L/5\cm)^{2}}{(W/0.3\cm)} \rm hr,~~~~~
\ena
which is a factor 2 smaller than the period estimated for the single size dust distribution (Eq. \ref{eq:Tcoll}).

Thus, the rotation by dust collisions is much less important than that by the equivalent electric field for $N_{\H}< 10^{17}\cm^{-2}$. When approaching $\alpha$ Centauri, the rotation by dust collisions is comparable to that by the equivalent electric field.

%\newpage
\section{Discussion}\label{sec:dis}

\subsection{Effect of spacecraft deflection}
We have found that the spacecraft can be deflected significantly by the interstellar magnetic field. For $\alpha$ Centauri ($D=1.34 \pc$), following Figure \ref{fig:dfl_uniform}, we find the deflection impact parameter is $b\sim 0.1\AU \sim 20R_{\odot}$ at $v=0.2c$ for $B_{\perp}\sim 3\mu$G) and $M_{\rm sp}=1\g$, assuming the maximum surface potential $U_{\rm max,e}$. {The deflection is 16 times larger when the maximum potential $U_{\rm max,\H}$ is adopted.} This deflection is significant since it amounts to twice the separation of Proxima b from its hosts star, Proxima Centauri.

The uncertainty in the deflection of the charged spacecraft by the interstellar magnetic field requires active course corrections by the spacecraft during its flight. These corrections can be achieved by thrusters.

\subsection{Effect of Spacecraft Oscillation and Rotation}
We have shown that the spacecraft can be rotated by the regular torque acting on the electric dipole moving through the magnetic field and impulsive torques from dust collisions. Such an effect can change the orientation of the spacecraft. 

In particular, the fast oscillation of the spacecraft can increase its frontal cross-sectioned area, which enhances the risks from dust impact, especially the chance of hitting very big grains that can completely evaporate the spacecraft with a single collision (\citealt{2017ApJ...837....5H}).

Also, the rotation could expose the spacecraft to enhanced gas bombardment, extending the charged region to the entire spacecraft volume.  As a result, the deflection by the magnetic field could be substantially increased by a factor of $LW^{2}/R_{\H}W^{2}\sim L/R_{\H}$.

It is worthy to mention that the rotation of a charged spacecraft can produce a magnetic moment, leading to the precession of the spacecraft around the ambient magnetic field.

Note that the dipole oscillation and rotation by dust bombardment should be treated simultaneously. Nevertheless, the dipole oscillation frequency is about one order of magnitude faster than the dust collision frequency. Moreover, a collision with a dust grain only deposits a small angular momentum to the spacecraft (see Section \ref{sec:rot}). Therefore, the separation of these two processes remains valid.

\subsection{Methods to mitigate spacecraft rotation}
To mitigate the instability due to dust collisions, we suggest launching spacecraft fast spinning around its long axis. The resulting large angular momentum can better align the spacecraft with the direction of motion.

To mitigate the rotation caused by the spacecraft charging, one can also put a shield of thickness $d$ satisfying $R_{e}< d < R_{\H}$. Such a shield can capture all impinging electrons and let most protons pass through. Since the electron yield is smaller than unity for $v>0.1c$ (see Figure \ref{fig:yield}), the shield will be negatively charged to the maximum potential of -10 kV. If the electron flow is collected, then, one can harvest an energy power of $\sim n_{\H}m_{e} v^{3}/2 \sim 2 {\rm mW}\cm^{-2}$ (B.T. Draine, private communication). Nevertheless, such a thin shield would be eroded by gas and dust collisions before the spacecraft reaches the target (\citealt{2017ApJ...837....5H}).

Another way to reduce the spacecraft charge is to launch the spacecraft with a net negative charge. It is also possible to use an electron gun to moderate the charge and acquire a better control on the spacecraft trajectory.

\subsection{Stellar corona and effect of plasma heating}
If entering the stellar corona as contemplated by \cite{2017ApJ...835L..32H}, the spacecraft will be rapidly heated to a high temperature by gas collisions \citep{2017ApJ...837....5H}. The typical density in active regions of solar corona declines with radius $R$ as $n_{i}\sim 10^{4}(100R_{\odot}/R)^{2}\cm^{-3}$ (\citealt{1999ApJ...523..812S}). Therefore, as shown in \cite{2017ApJ...837....5H}, the spacecraft will be melted for $R<100R_{\odot}$. We note that the {\it Parker Solar Probe} is intended to approach $\sim 10R_{\odot}$ to study the Sun's chromosphere and corona. However, protecting the Solar probe requires a shield of $11.3$cm thickness, which is too massive for gram-scale nanocrafts. 

\section{Summary}\label{sec:sum}

We studied the charging of the relativistic spacecraft by secondary electron emission due to collisions with interstellar gas. We have found that the front surface of the spacecraft will be rapidly charged to a maximum electric potential (charge) after it has swept a gas column of $N_{\H,\max}\sim 10^{15}\cm^{-2}$, on a scale much shorter than the distance to $\alpha$ Centauri.

We quantified the deflection of the charged spacecraft from a straight trajectory by the interstellar magnetic field. We found that the deflection distance can be larger than $20 R_{\odot}$, depending on the speed and magnetic field strength.

We also studied the oscillation of the spacecraft due to the interaction of the moving electric dipole with the magnetic field and the rotation by dust bombardment. We found that the electric dipole interaction can induce fast oscillation of the spacecraft around the axis perpendicular to the direction of motion, with a period of $\sim$ 0.5 hr. Impulsive torques by dust bombardment can also spin-up the spacecraft along its short axes, with a period of $\sim $1 hr.

Finally, we discussed the consequence of the spacecraft oscillation and rotation and suggested methods to mitigate its impact.\\

\acknowledgments
{We thank an anonymous referee for insightful comments.} We are grateful to B. Burkhart, A. Lazarian and Z. Manchester for insightful comments on the manuscript. T.H. thanks B.T. Draine for stimulating discussions on spacecraft charging. This work was supported in part by a grant from the Breakthrough Prize Foundation.

%\appendix

%--------------adding references-----------------------------------
%\bibliographystyle{/Users/thiemhoang/Dropbox/Papers2/apj}% or other styles: mcbride,plain, abbrv, acm, alpha, apalike, apj
%\bibliography{/Users/thiemhoang/Dropbox/Papers2/cites_paperApJ,/Users/thiemhoang/Dropbox/Papers2/cites_Books}
%\bibliographystyle{/home/home1/cthoang/Dropbox/Papers2/apj}
%\bibliography{/home/home1/cthoang/Dropbox/Papers2/cites_paperApJ,/home/home1/cthoang/Dropbox/Papers2/cites_Books}
\bibliography{ms.bbl}
\end{document}